\documentclass[floatfix,aps,twocolumn,prl,superscriptaddress, ]{revtex4}
\usepackage{amsmath}
\usepackage{amssymb}
\usepackage{graphicx}
\usepackage{dcolumn}
\usepackage{natbib}
\usepackage{bm}
\usepackage{epsfig}

\begin{document}

\title{Light quasiparticles dominate electronic transport in molecular
crystal field-effect transistors}
\author{Z.Q. Li}
\email{zhiqiang@physics.ucsd.edu}
\affiliation{Department of Physics, University of California, San Diego, La Jolla,
California 92093, USA}
\author{V. Podzorov}
\affiliation{Department of Physics and Astronomy, Rutgers University, Piscataway, New
Jersey 08854, USA}
\author{N. Sai}
\affiliation{Department of Physics, University of California, San Diego, La Jolla,
California 92093, USA}
\affiliation{Department of Physics, The University of Texas at Austin, Austin, Texas
78712, USA}
\author{M.C. Martin}
\affiliation{Advanced Light Source Division, Lawrence Berkeley National Laboratory,
Berkeley, California~94720, USA}
\author{M.E. Gershenson}
\affiliation{Department of Physics and Astronomy, Rutgers University, Piscataway, New
Jersey 08854, USA}
\author{M. Di Ventra}
\affiliation{Department of Physics, University of California, San Diego, La Jolla,
California 92093, USA}
\author{D.N. Basov}
\affiliation{Department of Physics, University of California, San Diego, La Jolla,
California 92093, USA}
\date{\today }

\begin{abstract}
We report on an infrared spectroscopy study of mobile holes in the
accumulation layer of organic field-effect transistors based on rubrene
single crystals. Our data indicate that both transport and infrared
properties of these transistors at room temperature are governed by light
quasiparticles in molecular orbital bands with the effective masses m$%
^{\star }$ comparable to free electron mass. Furthermore, the m$^{\star }$
values inferred from our experiments are in agreement with those determined
from band structure calculations. These findings reveal no evidence for
prominent polaronic effects, which is at variance with the common beliefs of
polaron formation in molecular solids.
\end{abstract}

\maketitle

A comprehensive understanding of charge transport in organic semiconductors
poses a significant intellectual challenge and, at the same time, is crucial
for further advances in the field of \textquotedblleft plastic
electronics\textquotedblright \cite{Malliaras,Forrest}. One longstanding
problem pertains to the nature of the electronic excitations responsible for
charge transport in these systems\cite{Capek,GershRMP}. A commonly used
description\cite{Capek} is that the electrical current in these easily
polarizable materials is carried by polarons: electrons or holes strongly
coupled to local lattice deformations. A hallmark of the polaronic transport
is a strong enhancement of the effective mass $m^{\star } $ compared to the
band values\cite{Capek}. Therefore, the hypothesis of polaron formation in
molecular crystals is verifiable since the effective masses of mobile
charges can be directly probed in infrared (IR) spectroscopic measurements.

Here we report on IR spectroscopy studies of charge carriers in the
conducting channel of organic field-effect transistors (OFET) based on
single crystals of rubrene (C$_{42}$H$_{28}$, Fig.1b inset), a
small-molecule organic semiconductor\cite{GershRMP,GerRev04,VPprl04,VPprl05}%
. These studies show that charge transport in rubrene based OFETs at room
temperature is dominated by light quasiparticles in the highest occupied
molecular orbital(HOMO) band. New spectroscopy data along with band
structure calculations help to elucidate recent observations of
non-activated, diffusive charge transport at the surface of high-quality
molecular crystals in studies of single-crystal OFETs\cite%
{GershRMP,GerRev04,VPprl04,VPprl05}.

\begin{figure}[tbp]
\includegraphics[width=6.766cm, height=7.688cm]{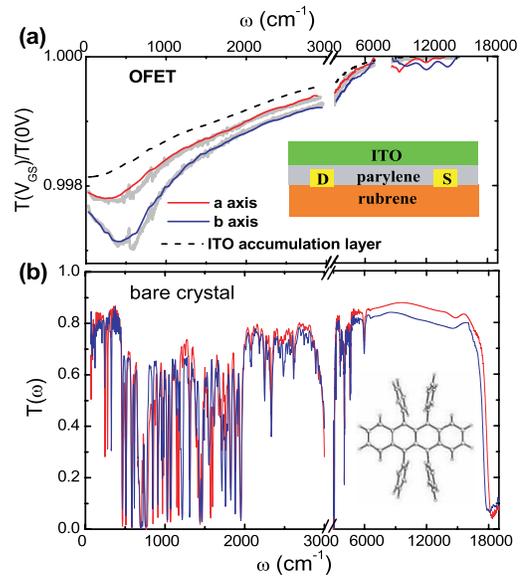}
\caption{(color online) (a): Voltage-induced changes of the transmission
spectra T($\protect\omega $,V$_{GS}$)/T($\protect\omega $,V$_{GS}$=0V) for a
representative rubrene OFET at V$_{GS}$=-280V at 300K. Thick gray lines:
model spectra at -280V obtained from a multilayer analysis as described in
the text. Dash line: the contribution of the ITO accumulation layer to the
raw T($\protect\omega $,V$_{GS}$)/T($\protect\omega $,V$_{GS}$=0V) spectra.
Inset of (a): A schematic of the OFETs. (b): The T($\protect\omega $)
spectra of a bare rubrene crystal. Inset of (b): The molecular structure of
rubrene.}
\label{Fig.1}
\end{figure}

A schematic of the rubrene OFETs studied here is displayed in the inset of
Fig. 1a. In these devices, source and drain graphite or silver paint
contacts were prepared on the surface of a rubrene single crystal followed
by the deposition of approximately 1 $\mu m$ of parylene that serves as the
gate insulator\cite{GerRev04}. The gate-channel capacitance per unit area $%
C_{t}$ in this type of devices is $\thicksim $2.1 nF/cm$^{2}$. For gate
electrode, we used a 24-nm-thick layer of InSnO$_{x}$ (ITO) with the
electron density $5\times10^{20}$ cm$^{-3}$ and the sheet resistance 300 $%
\Omega $/square, deposited by dc-magnetron sputtering in pure argon. The
gate electrode covers the entire device area (up to $3\times3 mm^{2}$). We
studied the IR response of numerous OFETs with typical DC transport mobility
5 cm$^{2}$V$^{-1}$s$^{-1}$ at room temperature. The use of the
semitransparent electrode ITO enabled spectroscopic studies of the
field-induced accumulation layer (AL) in rubrene from far-IR up to 2.2 eV,
the band gap of rubrene. In our IR measurements, free-standing rubrene OFETs
were illuminated with linearly polarized light over the frequency range 30 -
18000 cm$^{-1}$ (4 meV-2.2 eV) with a spectral resolution of 4 cm$^{-1}$
using a home-built set-up for broad band micro-spectroscopy. We investigated
the IR transmission of the OFETs, T($\omega $,V$_{GS}$), as a function of
frequency $\omega $ and the voltage applied between the gate and source
electrodes, V$_{GS}$\cite{Dressel}. The source and drain electrodes were
held at the same potential in most measurements.

We first examine the transmission spectra T($\omega $) of a bare rubrene
single crystal (Fig.1b), which are instructive for the understanding of the
response of rubrene OFETs. The sharp absorption lines below 5000 cm$^{-1}$
originate from phonons. The abrupt suppression of T($\omega $) at about
18000 cm$^{-1}$ is due to the lowest interband transition between HOMO and
LUMO (lowest unoccupied molecular orbital) bands. The interband transitions,
the phonon frequencies and the overall transmission level are all
anisotropic within the $ab$-plane.

Fig.1a depicts the T($\omega $,V$_{GS}$)/T($\omega $,V$_{GS}$=0V) spectra
for a representative OFET at 300K, obtained with the \textbf{E} vector along
the $a$ axis and $b$ axis. Similar results were found in all devices we
investigated. Under an applied gate voltage, a pronounced suppression of the
transmission of the transistor is observed, which is stronger for the
polarization of the E vector along the $b$-axis of the rubrene crystal. The
effect is most prominent at far-IR frequencies and peaked near 400 cm$^{-1}$%
. The form of T(V$_{GS}$)/T(0V) traces does not change appreciably with the
applied voltage, whereas the magnitude varies nearly linearly with V$_{GS}$.
These observations suggest that the changes of IR properties of our devices
are intimately related to the formation of ALs both in the channel of the
OFET and in the ITO gate electrode. Using IR microscopy\cite{LiNanoLett} we
have verified that the density of induced carriers is uniform along the
channel of the OFETs. Therefore, the two dimensional (2D) carrier density $%
n_{2D}$ in our OFETs can be estimated as follows: 
\begin{equation}
en_{2D}=C_{t}V_{GS}
\end{equation}%
where $e$ is the elementary charge.

\begin{figure}[tbp]
\includegraphics[width=7.101cm, height=7.071cm]{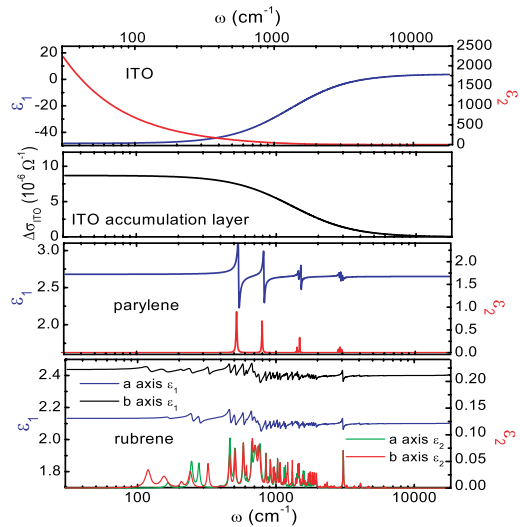}
\caption{(color online) Optical constants of all the constituent layers in
rubrene OFETs including the conductivity of the ITO accumulation layer $%
\Delta \protect\sigma _{ITO}(\protect\omega )$ at V$_{GS}$=-280V.}
\label{Fig.2}
\end{figure}

In order to characterize the dynamical properties of the charges in the
field-induced AL at the rubrene-parylene interface, it is imperative to
extract the optical constants of this layer. We employed an analysis
protocol that takes into account all the layers in our devices: 1) ITO, 2)
AL at the ITO-parylene interface characterized by the 2D conductivity $%
\Delta \sigma _{ITO}(\omega )$, 3) parylene, 4) AL at the rubrene-parylene
interface with the 2D conductivity $\Delta \sigma _{rub}(\omega )$ and 5)
bulk rubrene crystal. The response of layers 1, 3 and 5 was assumed to be
voltage independent, whereas layers 2 and 4 reveal voltage-dependent
properties. We first evaluated the complex dielectric function $\hat{%
\varepsilon}$($\omega $)=$\varepsilon _{1}$($\omega $)+i$\varepsilon _{2}$($%
\omega $)=1+4$\pi $i$\hat{\sigma}$($\omega $)/$\omega $ for layers 1, 3
and 5 from a combination of reflection, transmission and ellipsometric
measurements performed on layers with the same thicknesses as the ones used
in the OFETs, as shown in Fig. 2. We then extracted $\Delta \sigma
_{ITO}(\omega )$ from the Drude model: $\Delta \sigma _{ITO}(\omega )=\frac{%
n_{2D}e^{2}}{m^{\star }}\frac{\gamma _{D}}{\gamma _{D}^{2}+\omega ^{2}}$,
where the relaxation rate $\gamma _{D}$ =1300 cm$^{-1}$ and effective mass $%
m^{\star }$=0.5$m_{e}$ were determined from the measurements of ITO films,
and carrier density $n_{2D}$ of ITO AL was obtained from Eq.(1). Finally, we
employed a multi-oscillator fitting procedure\cite{Kuzmenko} to account for
the contribution of $\Delta \sigma _{rub}(\omega )$ to the transmission of
the device calculated from standard methods for multi-layered structures.
The $\Delta \sigma _{rub}(\omega )$ spectra extracted using this routine are
shown in Fig. 3 whereas the T(V$_{GS}$=-280V)/T(0V) spectra generated from
the multilayer model are plotted in Fig.1a. Alternatively the response of
our devices can be described by Eq.(3) in Ref.\cite{Tsui}, which directly
relates the raw T(V$_{GS}$)/T(0V) data to the 2D conductivity of the ALs.
The $\Delta \sigma _{rub}(\omega )$ spectra inferred from these two methods
are in agreement within an accuracy of 10 \%, which is comparable to the
error in the raw data. Fig.1a depicts the contribution of $\Delta \sigma
_{ITO}(\omega )$ to the raw T(V$_{GS}$)/T(0V) data obtained by omitting the
rubrene AL from the model, which shows that voltage-induced changes of the
electron density in ITO result in a measurable change in the transmission.
Note that the contribution of ITO to the T(V$_{GS}$)/T(0V) spectra has a
monotonic frequency dependence in contrast to the non-monotonic form of the
overall transmission change of the device.

\begin{figure}[tbp]
\includegraphics[width=5.034cm, height=6.217cm]{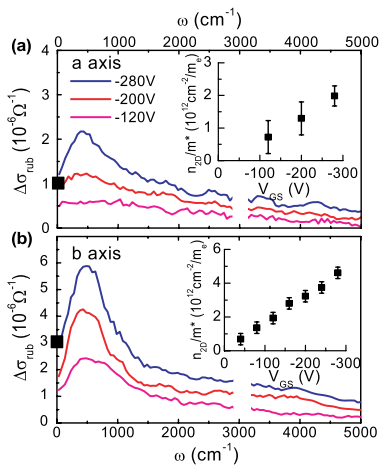}
\caption{(color online) The optical conductivity of the two-dimensional
system of field-induced charges at the rubrene-parylene interface $\Delta 
\protect\sigma _{rub}(\protect\omega )$ at different gate voltages V$_{GS}$
at 300K. (a): $E\parallel a$-axis data. (b): $E\parallel b$-axis data. Black
squares on the left axes: DC conductivity at -280V. Insets: The evolution of
the spectral weight $\frac{n_{2D}}{m^{\star }}$\ with V$_{GS}$.}
\label{Fig.3}
\end{figure}

Fig. 3 displays the optical conductivity spectra of the 2D system of
field-induced quasiparticles at the rubrene-parylene interface $\Delta
\sigma _{rub}(\omega )$ for the polarization of the \textbf{E} vector along
the $a$ and $b$ axes. Similar to the raw T(V$_{GS}$)/T(0V) data, the
conductivity spectra are characterized by a finite energy peak centered at
around 400 cm$^{-1}$. At lower frequencies, $\Delta \sigma _{rub}(\omega )$\
decreases toward the DC value that was obtained independently in DC
transport measurements. In the near-IR range, $\Delta \sigma _{rub}(\omega )$%
\ is negligibly small as shown in Fig.1a. In particular, no noticeable
features were observed in $\Delta \sigma _{rub}(\omega )$ at frequencies
close to the band gap of rubrene. The anisotropy of the conductivity spectra
is found throughout the IR range and extends to the DC limit. Importantly, $%
\Delta \sigma _{rub}(\omega )$ remains finite with the temperature
decreasing down to 30 K (not shown) in the entire IR range down to at least
80-90 cm$^{-1}$. We therefore conclude that no sizable energy gap opens up
in the IR response of the AL formed by voltage-induced holes in rubrene.

The non-monotonic form of the conductivity spectra in Fig. 3 can be
qualitatively described by the localization-modified Drude model that is
commonly used to account for the IR properties of organic materials and
other disordered conductors in the vicinity of metal-insulator transitions%
\cite{MottKaveh}. This description is not unique. We therefore will focus on
the overall strength of the $\Delta \sigma _{rub}(\omega )$ spectra, which
allows us to evaluate the optical effective mass $m^{\star }$ of the
field-induced quasiparticles using the model-independent oscillator strength
sum rule\cite{Wooten}, 
\begin{equation}
\frac{n_{2D}}{m^{\star }}=\int_{0}^{\Omega _{c}}\Delta \sigma _{rub}(\omega
)d\omega
\end{equation}%
The cutoff frequency $\Omega _{c}$ is chosen to be 5000 cm$^{-1}$ to
accommodate the entire energy region where voltage-induced changes are
prominent. $\frac{n_{2D}}{m^{\star }}$ increases linearly with V$_{GS}$ for
both $E\parallel a$ and $E\parallel b$ data (insets in Fig. 3), which is
consistent with the capacitive model Eq.(1) provided that $m^{\star }$ does
not change within the range of applied biases. This agreement further
justifies the use of Eq.(1) for extracting the charge density in the AL,
which approaches $3.7\times10^{12}$ cm$^{-2}$ at -280 V. The slopes of $%
\frac{n_{2D}}{m^{\star }}$(V$_{GS}$) for the $E\parallel a$ and $E\parallel
b $ measurements are different as shown in Fig. 3. This result in
conjunction with Eqs. (1), (2) yields the anisotropy of the effective mass: $%
m_{a}^{\star }=1.9\pm 0.3m_{e}$\ and $m_{b}^{\star }=0.8\pm 0.1m_{e}$. The
direct spectroscopic observation of the mass anisotropy elucidates the
origin of the anisotropic electronic mobility $\mu =e\tau /m^{\star }$
observed in transport measurements\cite{GershRMP,GerRev04,VPprl04,VPprl05}.
Indeed the magnitude of the anisotropy of $m^{\star }$ is in good agreement
with that of the mobility within experimental errors; this suggests that the
anisotropic effective masses of mobile quasiparticles dominate the
directional dependence of transport properties.

To understand the experimental data, we have carried out first-principles
density functional theory (DFT) calculations of the band structure of
rubrene within the generalized gradient approximation (GGA)\cite{Becke}
using the experimental lattice parameters\cite{Kafer}, as displayed in Fig.
4. The orthorhombic unit cell has two $ab$ planes separated by half a unit
cell in the $c$ direction, each containing two inequivalent rubrene
molecules. We find however that the interactions between the $ab$ planes in
the unit cell are negligible. From fitting the top of the HOMO band with
parabolic functions over different energy ranges along the $\Gamma $-X and $%
\Gamma $-Y directions, we obtained effective masses in the range of values
1.9$m_{e}<m_{a}^{\star }<$2.8$m_{e}$ and 1.0$m_{e}<m_{b}^{\star }<$1.3$m_{e}$
for hole carriers along the a and b directions, respectively. The DFT-GGA
functional may also contribute to errors in the computed band masses.
Nevertheless, the band mass values are overall comparable to those of the
field-induced quasiparticles inferred from IR measurements. The difference
between the two could be due to the errors in these quantities as discussed
above. Therefore, our IR study, along with band structure calculations,
reveals no significant enhancement of the effective mass of the
quasiparticles in rubrene OFETs compared to the band values.

\begin{figure}[tbp]
\includegraphics[width=5.324cm, height=4.161cm]{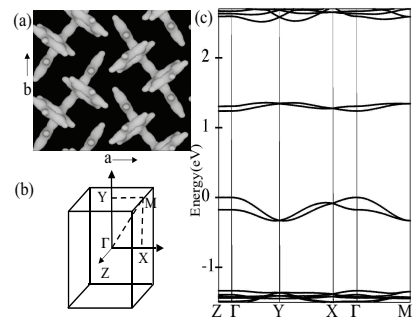}
\caption{(a): The charge density isosurface enclosing 40\% of the total
charges obtained from DFT, which illustrates the crystal structure in the $%
ab $ plane with two inequivalent rubrene molecules arranged in a herringbone
structure. (b): The reciprocal lattice; $\Gamma $-X, $\Gamma $-Y and $\Gamma 
$-Z correspond to the $a$, $b$ and $c$ crystalline direction. (c): The band
structure of rubrene calculated using DFT-GGA.}
\label{Fig.4}
\end{figure}

Our IR studies of charge dynamics in rubrene OFETs and band structure
calculations unveil several unexpected aspects of the quasiparticles in
these systems. First, no low-energy gap in the optical conductivity $\Delta
\sigma _{rub}(\omega )$ is observed in the whole IR range suggesting that
the field-induced quasiparticles reside in a continuum of electronic states
extending both above and below the Fermi energy. The states involved in
quasiparticle dynamics reflect the intrinsic electronic structure of rubrene
with rather distinct values of overlap integrals in the $a$- and $b$%
-directions. This conclusion is attested by the anisotropy of transport
properties, IR conductivity and most importantly by the effective masses
directly determined from IR data. Rather light effective masses of mobile
quasiparticles once again point to the involvement of band states (within
the HOMO band) in the electronic response in agreement with the band
structure analysis. These findings suggest that the periodic potential of
the molecular crystal lattice and the electronic band structure play a
dominant role in charge dynamics even at room temperature. Because organic
molecular crystals are periodic systems, the concept of energy bands in
these systems at sufficiently low temperatures is not in dispute. However,
in view of the weak inter-molecular van-der-Waals bonds in these crystals,
the long-range order may be disrupted by the thermally-induced dynamic
disorder\cite{Troisi}. Nevertheless, our results show that the band
dispersion evaluated in the limit of T$\rightarrow $0 provides an accurate
account of transport and IR properties at room temperature. The notion of
light quasiparticles in the HOMO band established through these findings is
furthermore supported by recent observations of non-activated, diffusive
charge transport on the surface of high-quality molecular crystals\cite%
{GershRMP,GerRev04,VPprl04,VPprl05}, also suggesting the existence of
extended electronic states.

Light effective masses comparable to band values reported here have not been
foreseen by theoretical models commonly postulating very strong coupling
between electronic and lattice degrees of freedom in molecular solids
leading to the formation of small polarons even at room temperature. Small
polarons are characterized by large masses of at least several times the
band mass due to the coupling with lattice\cite{Capek} in stark contrast
with our observations. Therefore, our work indicates that polaronic effects
in rubrene OFETs are weaker at room temperature than previously thought, and
the charge transport can be adequately described by quasiparticles in the
HOMO band. This assertion is furthermore supported by the frequency
dependence of the optical conductivity. Polarons in organic systems
(including OFETs) typically give rise to broad resonances in the absorption
spectra in mid-IR frequencies\cite{LiNanoLett} that are not detected in our
data for rubrene-based transistors. We conclude that the polaron binding
energies in rubrene must be below 26 meV, the energy that corresponds to
room temperature. Future work will be aimed at establishing if polaronic
effects in general, and enhancement of the effective mass in particular, may
be responsible for a rapid suppression of the conductivity below 140 K\cite%
{VPprl04}.

Work at UCSD is supported by the NSF, DOE and PRF. Work at Rutgers
University is supported by the NSF grants DMR-0405208 and ECS-0437932. The
Advanced Light Source is supported by the Director, Office of Science,
Office of Basic Energy Sciences, of the U.S. Department of Energy under
Contract No. DE-AC02-05CH11231.


\begin{thebibliography}{99}
\bibitem{Malliaras} G. Malliaras and R.H. Friend, Phys. Today 58, 53 (2005).

\bibitem{Forrest} S.R. Forrest, Nature 428, 911 (2004).

\bibitem{Capek} E.A. Silinsh and V. Capek, \textit{Organic Molecular
Crystals: Interaction, Localization, and Transport Phenomena} (AIP Press,
New York, 1994).

\bibitem{GershRMP} M.E. Gershenson, V. Podzorov, and A.F. Morpurgo, Rev.
Mod. Phys. 78, 973 (2006); C.H. Ahn et al., Rev. Mod. Phys. 78, 1185 (2006).

\bibitem{GerRev04} R.W.I. de Boer, M.E. Gershenson, A.F. Morpurgo, and V.
Podzorov, Phys. Status Solidi A 201, 1302 (2004).

\bibitem{VPprl04} V. Podzorov et al., Phys. Rev. Lett. 93, 086602 (2004).

\bibitem{VPprl05} V. Podzorov, E. Menard, J. A. Rogers, and M. E.
Gershenson, Phys. Rev. Lett. 95, 226601 (2005).

\bibitem{Dressel} IR properties of rubrene OFETs were also studied in
another recent work: M. Fischer et al., Appl. Phys. Lett. 89, 182103 (2006).

\bibitem{LiNanoLett} Z. Q. Li et al., Nano Lett. 6 , 224 (2006).

\bibitem{Kuzmenko} A. B. Kuzmenko, Rev. Sci. Instrum. 76, 083108 (2005); N.
Sai et al., Phys. Rev. B 75, 045307 (2007).

\bibitem{Tsui} D.C. Tsui et al., Surface Science 73, 419 (1978).

\bibitem{MottKaveh} N.F. Mott and M. Kaveh, Adv. Phys. 34, 329 (1985).

\bibitem{Wooten} F. Wooten, \textit{Optical Properties of Solids} (Academic,
New York/London, 1972).

\bibitem{Becke} A.D. Becke, Phys. Rev. A 38, 3098 (1988); C. Lee, W. Yang,
and R.G. Parr, Phys. Rev. B 37, 785 (1988). We used the $socorro$ package
for the calculations. (http://dft.sandia.gov/Socorro/mainpage.html). Common
density functionals such as the GGA used in this work do not account for van
der Waals forces, which determine the correct distance between molecules.
However, when the bonding distances are fixed at the experimental ones, this
deficiency has negligible effect on the electronic dispersion that is mainly
determined by the overlap of the electronic wave functions.

\bibitem{Kafer} D. Kafer and G. Witte, Phys. Chem. Chem. Phys. 7, 2850
(2005); E. Menard et al., Adv. Mater. 18, 1552 (2006).

\bibitem{Troisi} A. Troisi and G. Orlandi, Phys. Rev. Lett. 96, 086601
(2006).
\end{thebibliography}
\end{document}